\documentclass[twocolumn,preprintnumbers,showpacs,pre,floatfix,superscriptaddress,amsmath,amssymb]{revtex4}
\usepackage{epsfig}
\usepackage{subfigure}
\usepackage{amsmath}
\usepackage{color}
\usepackage{amssymb}
\usepackage{setspace}
\usepackage{graphicx}
\usepackage{dcolumn}
\usepackage{bm}
\usepackage{times}
\usepackage{enumerate}
\input epsf

\graphicspath{{figures/}}

\begin{document}

\author{Daniel B. Larremore}
\email{daniel.larremore@colorado.edu} 
\affiliation{Department of Applied Mathematics, University of Colorado at Boulder, Colorado 80309, USA}

\author{Marshall Y. Carpenter}
\affiliation{Department of Applied Mathematics, University of Colorado at Boulder, Colorado 80309, USA}

\author{Edward Ott}
\affiliation{Institute for Research in Electronics and Applied Physics, University of Maryland, College Park, Maryland 20742, USA}
\affiliation{Department of Physics and Department of Electrical and Computer Engineering, University of Maryland, College Park, Maryland 20742, USA}

\author{Juan G. Restrepo} 
\affiliation{Department of Applied Mathematics, University of Colorado at Boulder, Colorado 80309, USA}

\title{Statistical Properties of Avalanches in Networks}


\begin{abstract}
We characterize the distributions of size and duration of avalanches propagating in complex networks. By an avalanche we mean the sequence of events initiated by the externally stimulated 'excitation' of a network node, which may, with some probability, then stimulate subsequent firings of the nodes to which it is connected, resulting in a cascade of firings. This type of process is relevant to a wide variety of situations, including neuroscience, cascading failures on electrical power grids, and epidemology. We find that the statistics of avalanches can be characterized in terms of the largest eigenvalue and corresponding eigenvector of an appropriate adjacency matrix which encodes the structure of the network. By using mean-field analyses, previous studies of avalanches in networks have not considered the effect of network structure on the distribution of size and duration of avalanches. Our results apply to individual networks (rather than network ensembles) and provide expressions for the distributions of size and duration of avalanches starting at particular nodes in the network.  These findings might find application in the analysis of branching processes in networks, such as cascading power grid failures and critical brain dynamics. In particular, our results show that some experimental signatures of critical brain dynamics (i.e., power-law distributions of size and duration of neuronal avalanches), are robust to complex underlying network topologies.
\end{abstract}


\maketitle

\section{Introduction}

In this paper we study the statistics of avalanches propagating in complex networks. The study of avalanches of activity in complex networks is relevant to diverse fields, including epidemiology \cite{Allard2009,Miller2009}, genealogy \cite{GW}, and neuroscience \cite{Petermann2009,Shew2009,Stewart2008,Kinouchi2006,Larremore2011-PRL,Larremore2011-Chaos,Tanaka2009,Beggs2003,Shew2011,Benayoun2010}. The simplest case of an avalanche corresponds to a branching process \cite{Harris1963,Athreya1972}, first studied by Galton and Watson \cite{GW}, which can be considered as an avalanche propagating in a tree network. Various generalizations to the case where avalanches propagate in a more general network have been considered recently \cite{Samuelsson2006,Gleeson2008,Benayoun2010,Hackett2011}, and related problems such as the distribution of cluster size in percolation models \cite{Newman2001,Cohen2002} and self-organized criticality in the ``sandpile'' model \cite{Lee2004} have been studied. In contrast to these previous studies, we develop a theory of avalanche size and duration on complex networks that, instead of using some form of mean field analysis, explicitly includes the network topology.  This approach allows for an analysis of avalanches starting from arbitrary nodes in the network and the effect of nontrivial network structure on the distribution of avalanche size and duration.

Our formalism in this paper is general, describing dynamics with applications to a wide variety of systems. Our results are correspondingly general, but they may be of particular interest to those investigating recent experimental observations of avalanches of neuronal bursting in the mammalian cortex. When a neuron fires, it stimulates other neurons which may subsequently fire. When this linked activity occurs in a cascade, it is called a {\it neuronal avalanche} (experimentally, neuronal avalanches are observed propagating in functional networks where each node represents a group of neurons). Recent experiments have studied neuronal avalanches of activity in the brains of awake monkeys \cite{Petermann2009}, anesthetized rats \cite{Gireesh2008}, slices of rat cortex \cite{Shew2009,Beggs2003}, and humans \cite{Poil2008}. These studies found that when the tissue is allowed to grow and operate undisturbed in homeostasis \cite{Stewart2008}, both the size and temporal duration of neuronal avalanches is distributed according to a power-law. In contrast, the application of drugs that selectively decrease the activity of inhibitory or excitatory neurons results in avalanches with different statistics \cite{Shew2009}. Based on these observations, it has been argued and demonstrated experimentally that many neuronal networks operate in a critical regime that leads to power-law avalanche distributions \cite{Shew2009,Beggs2003,Gireesh2008,Poil2008}, maximized dynamic range \cite{Shew2009, Larremore2011-PRL, Larremore2011-Chaos, Kinouchi2006}, and maximized information capacity \cite{Shew2011, Tanaka2009, Beggs2003}. Therefore, it is of great interest to characterize this critical state and to understand how experimental signatures of criticality may change upon modification of the underlying network (e.g., changes induced by the drugs used in experiments).

We find that the statistical properties of avalanches are determined by spectral properties of the matrix whose entries $A_{mn}$ are the probabilities that the avalanche propagates from node $n$ to node $m$. In particular, the eigenvalue $\lambda$ of maximum magnitude (by the Perron-Frobenius theorem $\lambda$ is real and positive if $A_{mn} > 0$) and its associated eigenvector play a prominent role in determining the functional form and the parameters for the statistical distribution of avalanche size and duration. While many of our findings have analogous results in classical Galton-Watson branching processes \cite{Harris1963,Athreya1972}, we emphasize that our analysis allows us to identify how changes in network structure affect the parameters of the statistical distributions of avalanche size and duration. Moreover, our theory allows us to obtain the statistics of avalanches starting at particular network nodes.

This paper is organized as follows. In Sec.~\ref{section-formulation} we describe our model for avalanche propagation in  networks. In Secs.~\ref{section-duration} and \ref{section-size} we analyze the statistics of avalanche duration and size. In Sec.~\ref{section-experiments} we validate our analysis through numerical experiments. Section~\ref{section-discussion} presents further discussion and conclusions.

\section{Formulation}\label{section-formulation}

To model the propagation of avalanches in a network, we consider a network of $N$ nodes labeled $m=1,2,...,N$. Each node $m$ has a state $\tilde{x}_{m}=0$ or $1$. We refer to $\tilde{x}_{m} = 0$ as the {\it resting} state and to $\tilde{x}_{m} = 1$ as the {\it excited} state. At discrete times $t=0,1,...,$ the states of the nodes $\tilde{x}_{m}^{t}$ are simultaneously updated as follows: (i) If node $m$ is in the resting state, $\tilde{x}_{m}^{t} = 0$, it can be excited by an excited node $n$, $\tilde{x}_{n}^{t} = 1$, with probability $0\leq A_{mn} <1$, so that $\tilde{x}_{m}^{t+1} = 1$. (ii) The nodes that are excited, $\tilde{x}_{n}^{t}=1$, will deterministically return to the resting state in the next time step, $\tilde{x}_{n}^{t+1} = 0$. We therefore describe a network of $N$ nodes with a $N \times N$ weighted network adjacency matrix $A = \{ A_{mn} \} $, where $A_{mn} > 0$ may be thought of as the strength of connection from node $n$ to node $m$, and $A_{mn} = 0$ implies that node $n$ does not connect to node $m$. We will assume that given any two nodes $n$ and $m$, the probability that an excitation originating at node $n$ is able to excite node $m$ (through potentially many intermediate nodes) is not zero. This is equivalent to saying the network is fully connected, and therefore the matrix $A$ is {\it irreducible}.

\begin{figure}[h]
	\centering
	\epsfig{file=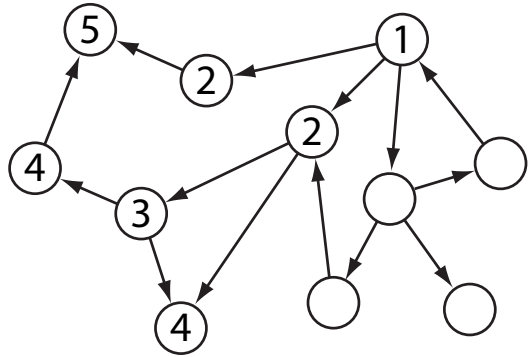, clip =,width=0.8\linewidth }
	\caption{An example avalanche is shown, where circles represent nodes, arrows represent links, and numbers inside nodes correspond to the time step at which each node is activated. Starting from a single excited node, labeled 1, the avalanche spreads to two other nodes, labeled 2, and so on. Note that the presence of a link does not guarantee the transmission of excitation. The example avalanche above lasts for five time steps and excited a total of six nodes in addition to the initial node, so $d=5$ and $x=7$.}
	\label{treeExample}
\end{figure}

Starting from a single excited node $k$ ($\tilde{x}_{n}^{0} = 1$ if $m = n$ and $\tilde{x}_{m}^{0} = 0$ if $m\neq n$), we let the system evolve according to the dynamics above, and observe the cascade of activity until there are no more excited nodes. This motivates the following definitions, which are illustrated in Fig. \ref{treeExample} : (1) an {\it avalanche} is the sequence of excitations produced by a single excited node; (2) the {\it duration} $d$ of an avalanche is defined as the total number of time steps spanned by the avalanche: if the avalanche starts with $\tilde{x}_{n}^0 = 1$, then 
\begin{align}
	\label{define-duration}
	d_{n} = \min_{t\geq 0}\{ \tilde{x}_{k}^t = 0\mbox{ for all } k \}.
\end{align}  
An avalanche that continues indefinitely  is said to have infinite duration; (3) the {\it size} $x$ of an avalanche starting at $\tilde{x}_{n}^{0}=1$ is defined as the total number of nodes excited during an avalanche, allowing for nodes to be excited multiple times: 
\begin{align}
	\label{define-size}
	x_{n} = \sum_{t=0}^{d-1} \sum_{k=1}^N\tilde{x}_{k}^t.
\end{align}
Note that it is possible for an avalanche to have size larger than the total size of the network (e.g., if $d_{n} = \infty$, then $x_{n} = \infty$). Our goal in this paper is to determine the probability distributions of these variables in terms of the matrix $A$.

\section{Distribution of Avalanche Duration}\label{section-duration}

In order to analyze the statistics of avalanche duration, we define $c_n(t)$ as the probability that an avalanche starting at node $n$ has duration less than or equal to $t$,
\begin{equation}\label{eq-definecn}
	c_{n}(t) = \text{P} ( d_n \leq t ).
\end{equation}
The quantity $c_n(t)$ is the cumulative distribution function (CDF) of the random variable $d_n$.  In what follows, we will restrict our attention to a class of networks that we call \emph{locally tree-like}. By locally tree-like, in this paper we shall mean that, for any given $t$ not too large, and pair of nodes $j$ and $k$, if there exists a directed path of length $t$ from $j$ to $k$, then it is rarely the case that there will also  exist a second such path \cite{Melnik2011}. Many networks found in applications are of this type, and it has been found that the locally tree-like approximation works very well in describing various dynamical processes while still capturing the effects of network heterogeneity \cite{Larremore2011-PRL, Larremore2011-Chaos, RestrepoOttHunt2006, Melnik2011,genes}. For these networks, we can approximately treat the avalanches propagating to different neighbors of node $n$ as independent, and write the recursion relation
\begin{equation}\label{eq-durationupdate}
c_n(t+1) = \prod_{m=1}^N \Bigl[(1-A_{mn}) + A_{mn}c_m(t) \Bigr],
\end{equation}
together with $c_{n}(0) = 0$ which follows from the definition \eqref{eq-definecn}. 
The right hand side of Eq.~\eqref{eq-durationupdate} is the probability that nodes are either not excited by node $n$, or, if they are, that they generate avalanches of duration at most $t$: $(1-A_{mn})$ is the probability that an excitation {\it does not} pass from node $n$ to node $m$, whereas $A_{mn} c_{m}(t)$ is the probability that an excitation {\it does} pass from node $n$ to node $m$ and the resulting avalanche has duration at most $t$.  Note that Eq.~\eqref{eq-durationupdate}  can treat any node $n$ as the starting node for an avalanche. As discussed above, Eq.~\eqref{eq-durationupdate} assumes that the descendent branches of the avalanche are independent.  It is, however, possible that an avalanche may branch in such a way that two branches interact at a later time. Nevertheless, for the networks we studied we found that, while these events do occur for large avalanches, they do not significantly affect our predictions. We show numerical results confirming this in Sec.~\ref{section-experiments}.

We are interested in the distribution of long avalanche duration, i.e., in the asymptotic form of $c_n(t)$ for $t\to \infty$. By definition (see also Appendix \ref{appendixa}), $c_n(t)$ is a bounded, increasing function of $t$, and therefore it must converge to a value $\lim_{t\to \infty} c_n(t) = b_n \leq 1$ which can be interpreted as the probability that an avalanche starting at node $n$ has finite duration. Our analysis will be based on whether or not this limit is strictly less than one or equal to one. As shown in Appendix \ref{appendixa}, this is determined by the Perron-Frobenius eigenvalue of $A$, $\lambda$: if $\lambda \leq 1$, then $\lim_{t\to \infty} c_n(t) = 1$. The case $\lambda < 1$ will be referred to as the {\it subcritical} case, and the case $\lambda = 1$ will be referred to as the {\it critical} case. On the other hand, if $\lambda > 1$, then $\lim_{t\to \infty} c_n(t) = b_n < 1$, which implies that there is a nonzero probability that an avalanche has infinite duration. This case will be referred to as the {\it supercritical case}. The asymptotic form of $c_n(t)$ will be analyzed separately for these three cases below.

\subsection{Subcritical Networks $(\lambda <  1)$}\label{subcritical}

In the subcritical case,  $b_n = 1$ is the only fixed point of the system Eq.~\eqref{eq-durationupdate} (see Appendix \ref{appendixa}). To analyze the asymptotic form of $c_n(t)$, we assume it is close to the fixed point and define the small quantity $f_n(t) = 1-c_n(t)$. Linearizing Eq.~\eqref{eq-durationupdate} we obtain
\begin{equation}
f_n(t+1) = \sum_{m=1}^N A_{mn}f_m(t).
\end{equation}
Assuming exponential decay (or growth) of perturbations, $f_n(t) = \lambda^{t}v_{n}$, we obtain
\begin{equation}\label{eq-eigenvalue}
\lambda v_n = \sum_{m=1}^N A_{mn}v_m .
\end{equation}
Thus, $\lambda$ is an eigenvalue of $A$ and ${\bf v} = [v_{1},v_{2},...,v_{N}]$ its left eigenvector. We identify $\lambda$ as the Perron-Frobenius eigenvalue since, having the largest magnitude among all the eigenvalues, $\lambda^t v_n$ will be the dominant term as $t\to \infty$ when compared with the other modes. We note that for finite $t$, this approximation is good as long as there is a large enough separation between $\lambda$ and the rest of the spectrum of $A$. This issue is discussed in \cite{Chauhan2009}, where it is found that this separation is typically large in networks without strong community structure. Henceforth, we will assume that $\lambda$ is well separated from the rest of the spectrum of $A$.
Therefore, $c_n(t)$ approaches $1$ exponentially as 
\begin{equation}
	\label{eq-expApproach1}
	c_n(t) \approx 1 - \lambda^t v_n,
\end{equation} 
where ${\bf v}$ is the left eigenvector of $A$ corresponding to  $\lambda$; $v_{n} > 0$ by the Perron-Frobenius theorem \cite{PFTheorem}. The fixed point $b_n = 1$ is linearly stable when $\lambda < 1$.

The probability density function (PDF) of avalanche duration  is given by $p_n(t) = P(d_n = t) = c_n(t) - c_n(t-1)$, so
\begin{align}\label{pdf1}
p_n(t) \sim (\lambda^{-1}-1) v_n \lambda^t ,
\end{align}
which decays exponentially to zero with decay rate $\ln(1/ \lambda)$.

In summary, we can draw two predictions from the analysis above for subcritical networks: (i) the PDF of avalanche duration decays exponentially towards zero as $\lambda^t$, and (ii) the probability that an avalanche started at node $n$ lasts  $t$ steps is proportional to the $n^{th}$ entry of the left eigenvector of $A$, $v_n$. These predictions are tested in Sec. \ref{section-experiments}.

\subsection{Supercritical networks $(\lambda >  1)$} \label{sec-supercriticalDuration}

A linear stability analysis of the fixed point $b_n = 1$ in the supercritical case shows that this fixed point is linearly unstable. This implies (see Appendix \ref{appendixa}) that there exists another fixed point $b_n$ to which $c_n(t)$ converges from below, $\lim_{t\to\infty}c_n(t) = b_n < 1$. Thus, there is a nonzero probability that an avalanche will have infinite duration. Our analysis below characterizes the distribution of finite avalanche duration in supercritical networks. We first note that the fixed point $b_n$ satisfies
\begin{equation}
	\label{eq-bnFixedPoint}
	b_n = \prod_{m=1}^N \Bigl[ (1-A_{mn}) + A_{mn}b_m \Bigr].
\end{equation}
Again, we introduce the quantity $f_n(t) = b_n -c_n(t)$, and consider the limit when $t$ is large and $f_n$ is small. We substitute this into Eq.~\eqref{eq-durationupdate} and rewrite it as
\begin{equation}\label{eq-halfwaytoD}
b_n -f_n(t+1) = b_n \prod_{m=1}^N \Bigl[ 1 - \frac{A_{mn}f_m(t)}{(1-A_{mn})+A_{mn}b_m} \Bigr].
\end{equation}
By defining a new matrix $D$ with entries
\begin{equation}
	\label{eq-supereigenvalue}
	D_{mn} = \frac{A_{mn}b_n}{(1-A_{mn})+A_{mn}b_m},
\end{equation}
and linearizing Eq.~\eqref{eq-halfwaytoD}  we find
\begin{equation}\label{eq-supercriticaldecay}
f_n(t+1) \approx  \sum_{m=1}^{N} D_{mn} f_m(t).
\end{equation}
As in the subcritical case, we conclude that $f_n(t) \approx \lambda_D^t w_n$, where ${\bf w}$ is the left Perron-Frobenius eigenvector of the matrix $D$
and $\lambda_D$ its Perron-Frobenius eigenvalue. As argued in Appendix~\ref{appendixb}, $\lambda_D <1$ when $\lambda > 1$, thus ensuring exponential convergence. Therefore, we have
\begin{equation}
	\label{eq-supercriticaldurationprediction}
	c_n(t) \approx b_n -  w_{n} \lambda_{D}^{t}.
\end{equation} 

As in the subcritical case, the probability density function (PDF) of avalanche duration  is given by
\begin{align}\label{pdf2}
p_n(t) \sim (\lambda_D^{-1}-1) w_n \lambda_D^t,
\end{align}
which decays exponentially to zero with decay rate $\ln(1/\lambda_D)$.

In summary, for supercritical networks: (i) the PDF of avalanche duration decays exponentially towards zero as $\lambda_D^t$, and (ii) the probability that an avalanche started at node $n$ lasts  $t$ steps is proportional to the $n^{th}$ entry of the left eigenvector of $D$, $w_n$. These predictions are tested in Sec. \ref{section-experiments}. We note that these predictions simplify to those drawn from Eq.~\eqref{pdf1} if the network is subcritical, in which case $b_n = 1$, Eq.~\eqref{eq-supereigenvalue} simplifies to $D_{mn}=A_{mn}$, and therefore $\lambda_{D} = \lambda$ and ${\bf w} = {\bf v}$.

\subsection{Critical Networks $(\lambda = 1)$}\label{section-criticalduration}

The analyses above show that if $\lambda = 1$, the fixed point $b_n = 1$ is marginally stable. This fixed point must be an attracting fixed point, since $c_n(t)$ is nondecreasing and $b_n =1$ is the only fixed point of Eq.~(\ref{eq-durationupdate}) as shown in Appendix \ref{appendixa}.  To determine the asymptotic form of $c_n(t)$ for large $t$, we let $c_n(t) = 1 - f_n(t)$. We assume that Eq.~\eqref{eq-durationupdate} has a solution whose asymptotic functional form in $t$ (to be determined) can be extended to a differentiable function of a continuous time variable $t$. Since the convergence of $f_n(t)$ to $0$ is slower than exponential, we look for a solution $f_n(t)$ which is  slowly varying in $t$ when $f_n(t)$ is small, and approximate 
\begin{equation}\label{eq-slowlyvarying}
	f_{n}(t+1) \approx f_{n}(t) + f_{n}'(t).
\end{equation}
The slowly varying assumption implies that $df_{n}(t)/dt \equiv f_n'(t) \ll f_n(t)$ as $f_n\to 0$, which we note excludes an exponential solution. Substituting Eq.~\eqref{eq-slowlyvarying} into Eq.~\eqref{eq-durationupdate}, we get 
\begin{align}
	1- f_n(t) - f_n'(t) \approx \prod_{m=1}^N\left[1-A_{mn}f_m(t)\right].
\end{align}
Assuming $f_n(t)\ll 1$ and expanding to second order, we get after simplifying and dropping the time notation for clarity,
\begin{equation}
\label{eq}
	f_n+f_n' \approx  \sum_mA_{mn}f_m - \frac{1}{2}\sum_m\sum_{k\neq m} A_{mn} A_{kn} f_m f_k.
\end{equation}
The leading order terms are $f_n$ on the left-hand side and $\sum_m A_{mn}f_m$ on the right-hand side, so for these to balance as $f\to 0$ requires
\begin{align}
	f_n =  \sum_mA_{mn}f_m.
\end{align}
Therefore, in this limit the vector ${\bf f}(t) = [f_1(t),f_2(t),\dots,f_N(t)]^T$ has to be proportional to the normalized left eigenvector ${\bf v}$ of $A$ with eigenvalue $\lambda = 1$. Thus, a slowly varying solution only exists for a critical network. Since ${\bf v}$ is independent of time, the constant of proportionality must be time dependent, $f_n(t) = K(t) v_n$. Now, for finite $f$, we expect the solution to deviate by a small error from this limit solution, so we set 
\begin{align}
f_n(t) = K(t)v_n/\langle v\rangle + \varepsilon_n(t),
\end{align}
where we assume $\varepsilon_n \ll f_n(t)$, $\varepsilon'_n \ll f'_n(t)$, and the term $\langle v\rangle =\sum_{n=1}^N v_n/N$ is included to make $K(t)$ independent of the normalization of ${\bf v}$. Inserting this in Eq.~\eqref{eq}, neglecting terms of order $\varepsilon'$, $\varepsilon^2$, $f\varepsilon$, and using $\sum_m\sum_{k\neq m} A_{mn} A_{kn} v_mv_k \approx  v_n^2$, we obtain
\begin{align}
 \varepsilon_n + K'(t)v_n/\langle v\rangle = \sum_{m=1}^N A_{mn} \varepsilon_m -\frac{1}{2} K^2(t)v_n^2/\langle v\rangle^2.
\end{align}

To eliminate the unknown error term $\varepsilon$, we multiply by $u_n$, where ${\bf u}$ is the right eigenvector of $A$ satisfying $A{\bf u}={\bf u}$, and sum over $n$. The error terms cancel and we obtain an ordinary differential equation (ODE),
\begin{align}
	\label{ode}
K'(t) =  -\frac{1}{2} \frac{\langle u v^2\rangle}{\langle u v \rangle\langle v\rangle} K^2(t),
\end{align}
where $\langle x y\rangle  \equiv \frac{1}{N}\sum_nx_ny_n$. Solving this ODE yields
\begin{equation}
	K(t) \approx \frac{1}{\beta + \frac{1}{2} \frac{\langle u v^2\rangle}{\langle u v \rangle\langle v\rangle} t},
\end{equation}
where $\beta$ is an integration constant. In terms of the original variables, we obtain
\begin{equation}
	\label{eq-minusone}
	c_n(t) \approx 1 -\frac{v_n}{\beta + \frac{1}{2} \frac{\langle u v^2\rangle}{\langle u v \rangle\langle v\rangle} t}.
\end{equation}
The PDF, in the continuous time approximation, is given by $p_n(t) = c_n'(t)$,
\begin{align}\label{eq-pdf}
p_n(t) \propto  \frac{v_n}{\left(\beta + \frac{1}{2} \frac{\langle u v^2\rangle}{\langle u v \rangle\langle v\rangle} t\right)^2}.
\end{align}

From Eq.~\eqref{eq-pdf} we make the prediction that as $t \to \infty$, $p_n(t) \sim v_{n}t^{-2}$. This prediction is tested in Sec. \ref{section-experiments}.

\section{Distribution of Avalanche Size}\label{section-size}

In order to analyze the distribution of avalanche size, we define the random variable  $x_n$ as the size of an avalanche starting at node $n$. Let $z_{mn}$ be a random variable which is $1$ if node $n$ excites node $m$ and $0$ otherwise, so that $z_{mn} = 1$ with probability $A_{mn}$ and $0$ with probability $1-A_{mn}$. Thus
\begin{equation}
	\label{eq-sizeprobability}
	x_n = 1 + \sum_{m=1}^{N} z_{mn} x_m.
\end{equation}
When $\lambda > 1$ there is a nonzero probability that an avalanche has infinite duration, and therefore infinite size, as demonstrated in Sec.~\ref{sec-supercriticalDuration} and Appendix~\ref{appendixa}. Therefore, we will restrict our attention only to the distribution of avalanches that are finite. To study this distribution, we define the moment generating function
\begin{equation}
	\label{mgf}
	\phi_n(s) \equiv E[e^{-sx_n}|x_n < \infty].
\end{equation}
We now use Eq.~\eqref{eq-sizeprobability} to derive a relation between the moment generating functions corresponding to different nodes. First, we rewrite the condition $x_n < \infty$ for node $n$ in terms of events applicable to its neighbors. An avalanche starting at node $n$ is finite if and only if for every node $m$, either (i) the excitation does not pass from node $n$ to node $m$, or (ii) the excitation passes from node $n$ to node $m$ but the subsequent avalanche starting from node $m$ is finite. Therefore, we rewrite the condition $x_n < \infty$ as the requirement that for any $m$, $(z_{mn}, x_m) \in Z_{mn} \cup W_{mn}$, where we have defined the disjoint sets of events $Z_{mn} = \{z_{mn}=0\}$ and $W_{mn} = \{x_m < \infty$ and $z_{mn}=1\}$.
Assuming the independence of the random variables $x_m$ (consistent with the locally tree-like assumption used in the previous section), we can rewrite $\phi_n(s)$ as 
\begin{align}
	\label{ifindependent}
	\phi_n(s) &= e^{-s}\prod_{m=1}^{N} E\Bigl[ e^{-sz_{mn}x_m} | Z_{mn} \cup W_{mn} \Bigr],
\end{align}
where the expectation $E[\cdot]$ is taken over realizations of the random pairs $(z_{mn}, x_m)$.
Denoting $P(W)$ as the probability of an event set $W$, we relate the expected value in the product in Eq.~\eqref{ifindependent} to the probabilities of the events $W_{mn}$ and $Z_{mn}$:
\begin{align}\label{mess}
	E&[e^{-sz_{mn}x_m}| Z_{mn}\cup W_{mn}]P(Z_{mn}\cup W_{mn}) \notag \\
	&=E[e^{-sz_{mn}x_m}|Z_{mn}]P(Z_{mn}) \notag\\
	&+ E[e^{-sz_{mn}x_m}|W_{mn}]P(W_{mn}).
\end{align}
Using the following relations that follow from the definitions above,
\begin{align}
P(Z_{mn}) &= 1-A_{mn},\nonumber\\
P(W_{mn}) &= A_{mn} b_m,\nonumber\\
P(Z_{mn}\cup W_{mn}) & = (1-A_{mn}) + A_{mn} b_m,\nonumber\\
E[e^{-sz_{mn}x_m}|W_{mn}] &= \phi_m(s),\nonumber\\
E[e^{-sz_{mn}x_m}|Z_{mn}] &= 1\nonumber,
\end{align}
Substitution into Eq.~\eqref{mess} gives
\begin{align}
	E[e^{-sz_{mn}x_m}|Z_{mn}\cup W_{mn}]\nonumber \\
	= \frac{(1-A_{mn})+b_mA_{mn}\phi_m(s)}{(1-A_{mn})+b_mA_{mn}}.
\end{align}
Inserting this into Eq.~\eqref{ifindependent} we obtain  one of our main results,
\begin{equation}
	\label{eq-mainresult}
	\phi_n(s) = e^{-s}\prod_{m=1}^{N} \frac{(1-A_{mn})+b_mA_{mn}\phi_m(s)}{(1-A_{mn})+b_mA_{mn}}.
\end{equation}
\begin{figure*}[t]
	\centering
	\epsfig{file=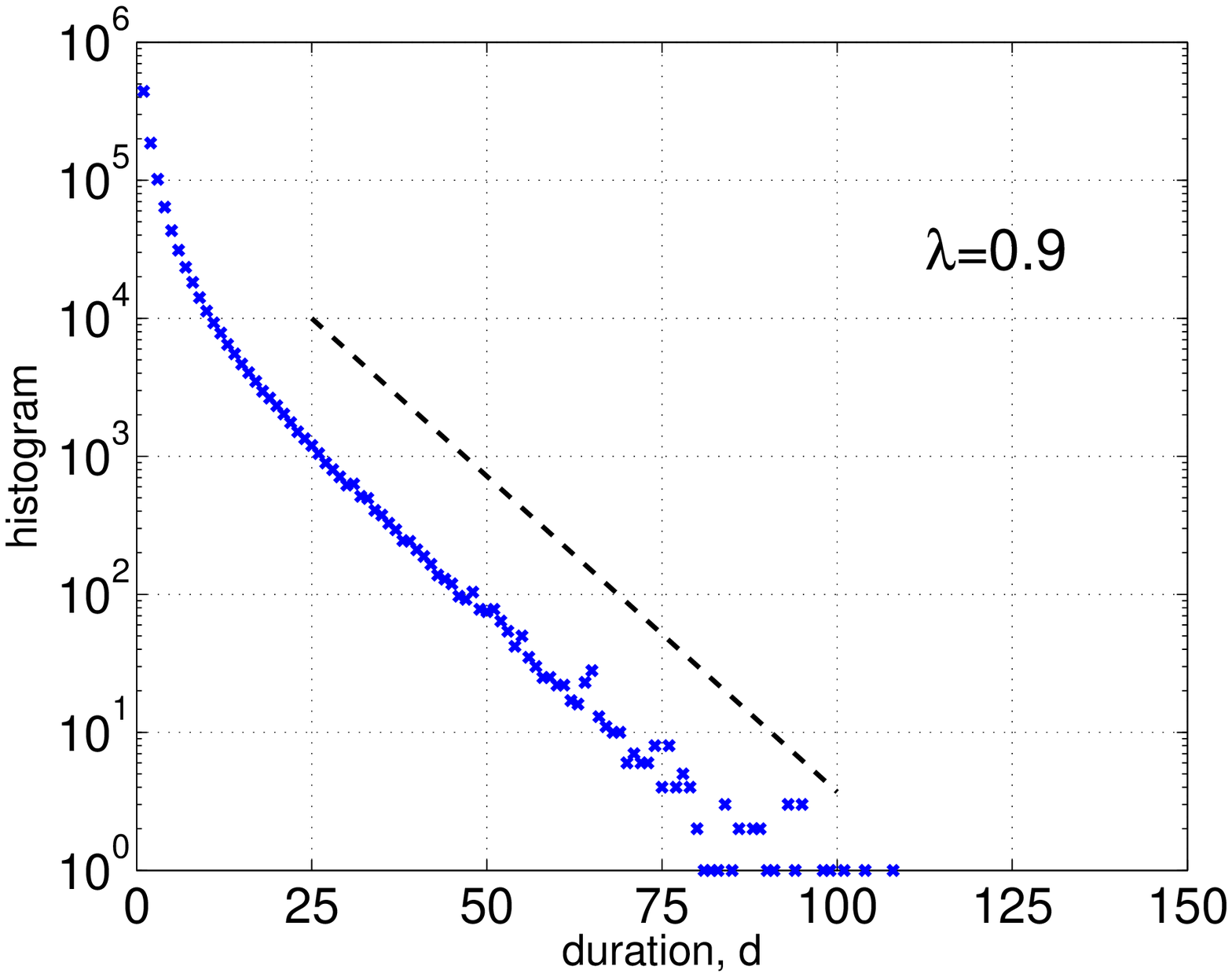, clip =,width=0.32\linewidth }
	\epsfig{file=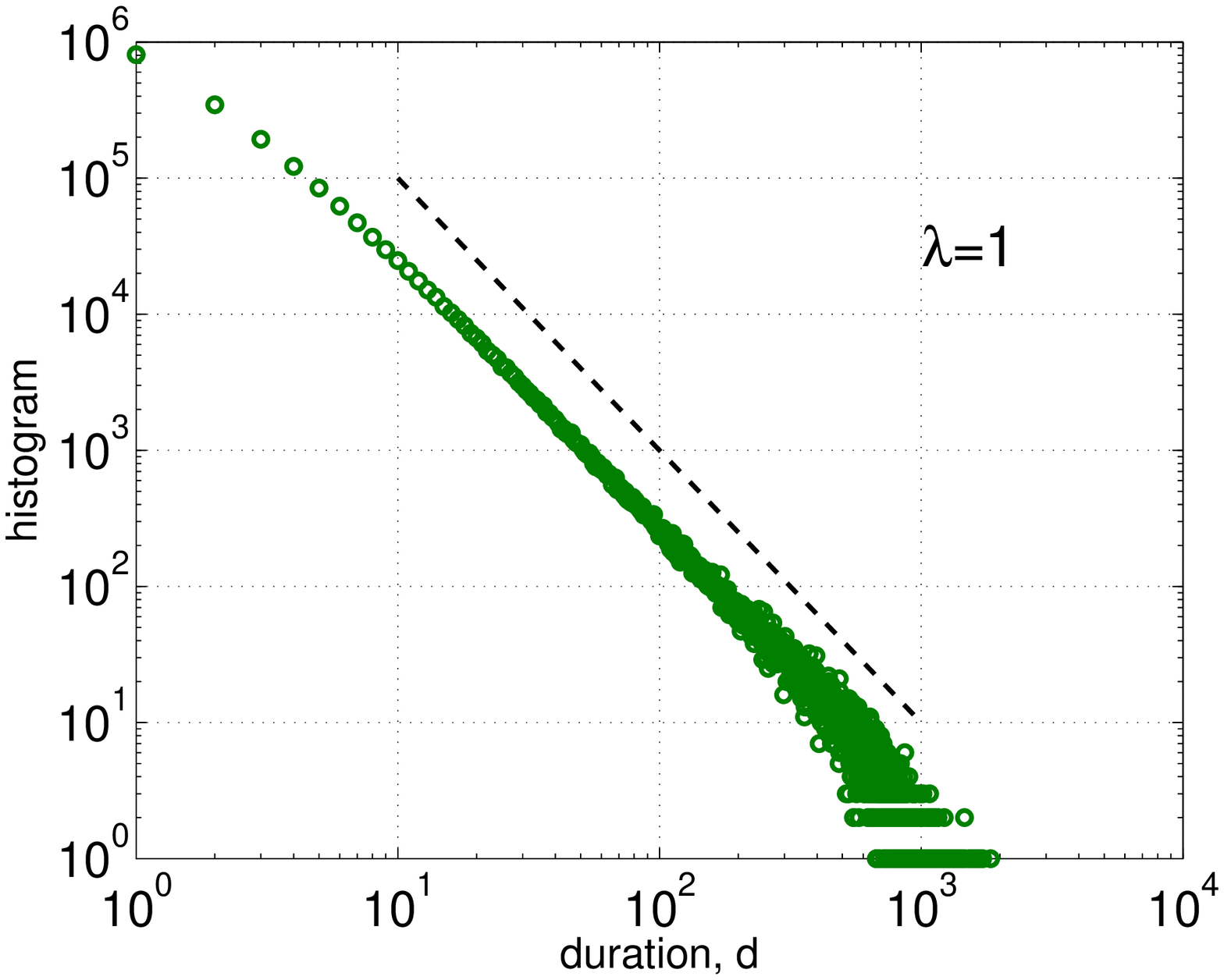, clip =,width=0.32\linewidth }
	\epsfig{file=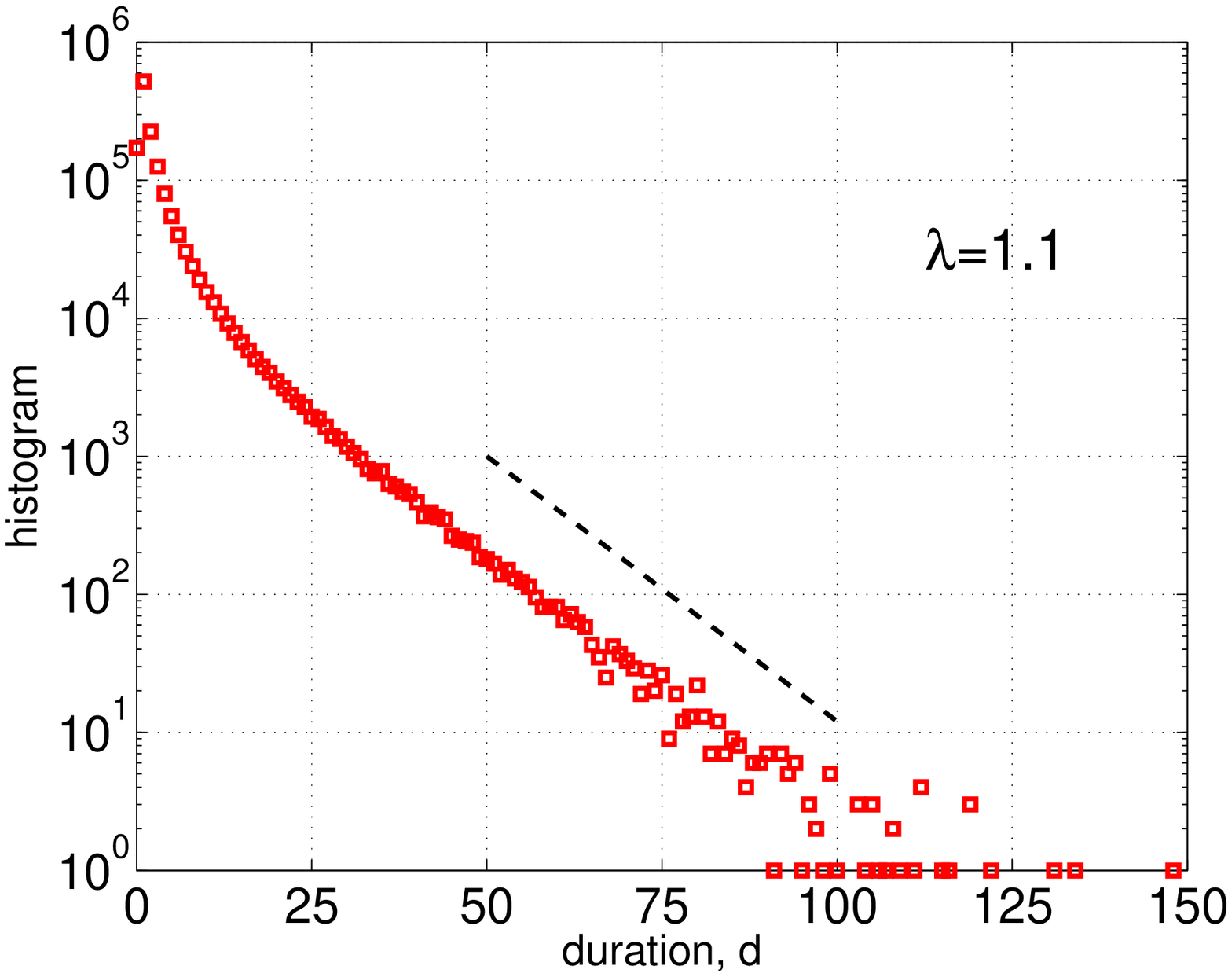, clip =,width=0.32\linewidth }
	\caption{(color online) Histograms of avalanche duration shown above for networks of $N=10^5$ nodes with power-law degree distribution, exponent $\gamma = 3.5$ with Perron-Frobenius eigenvalues of $\lambda = 0.9$ (left), $\lambda=1.0$ (center) and $\lambda=1.1$ (right). Symbols show the number of avalanches having duration $d$ from a single simulation of $10^6$, $2 \times 10^6$, and $10^6$ avalanches, respectively, from left to right. Dashed lines provide a reference for the theoretical predictions described in Eqs.~\eqref{eq-expApproach1}, \eqref{eq-minusone}, and \eqref{eq-supercriticaldurationprediction}. Note that the vertical position of the dashed lines was chosen arbitrarily. Due to predictions of exponential decay for the sub- and super-critical cases, the left and right plots are plotted on a log-linear scale, while the center plot is plotted on a log-log scale to show the power-law decay. Infinite duration avalanches in the supercritical case (right) are not displayed in the figure.}
	\label{figure-duration}
\end{figure*}
Defining $g_n(s) = \phi_n(s) - 1$, and the matrix $H$ with entries
\begin{equation}\label{eq-h}
	H_{mn} = \frac{b_mA_{mn}}{(1-A_{mn})+b_mA_{mn}}.
\end{equation}
we can rewrite Eq.~\eqref{eq-mainresult} as
\begin{equation}
	\label{eq-gn}
	1+g_n(s) = e^{-s}\prod_{m=1}^{N} [1+H_{mn}g_m(s)].
\end{equation}
Defining the $N \times N$ matrix, $B = \text{diag}(b_{1},b_{2},...,b_{N})$, we have from Eqs.~\eqref{eq-supereigenvalue} and \eqref{eq-h}, that $HB^{-1} = B^{-1}D$. Thus the matrix $H$ is related to the matrix $D$ by a similarity transformation and thus has the same spectrum. Therefore, we will denote the Perron-Frobenius eigenvalue of $H$ by $\lambda_D$. Note that $\lambda_D = \lambda$ when $\lambda \leq 1$, since in that case $b_n = 1$ and $H = A$.
\begin{figure*}[t]
	\centering
	\epsfig{file=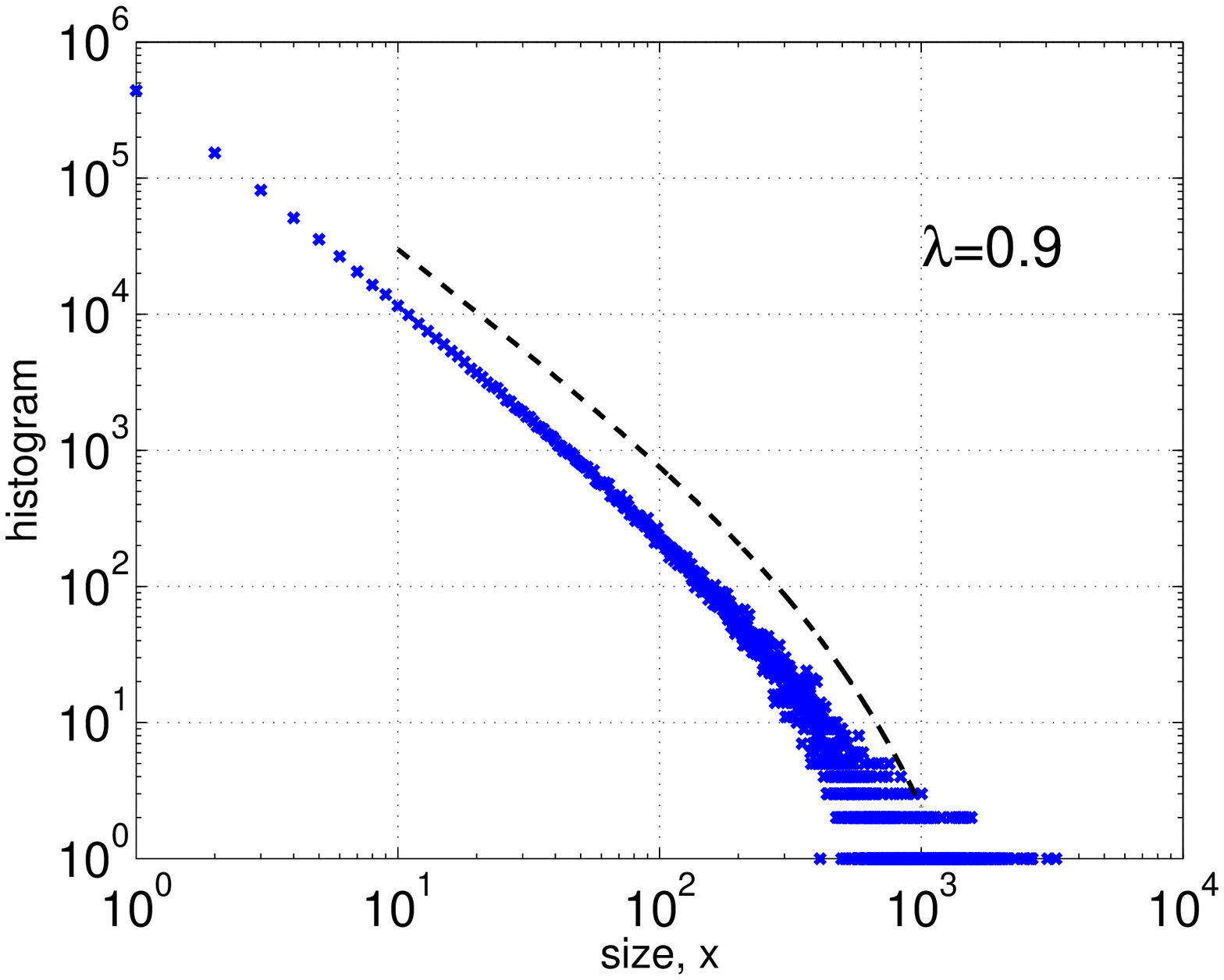, clip =,width=0.32\linewidth }
	\epsfig{file=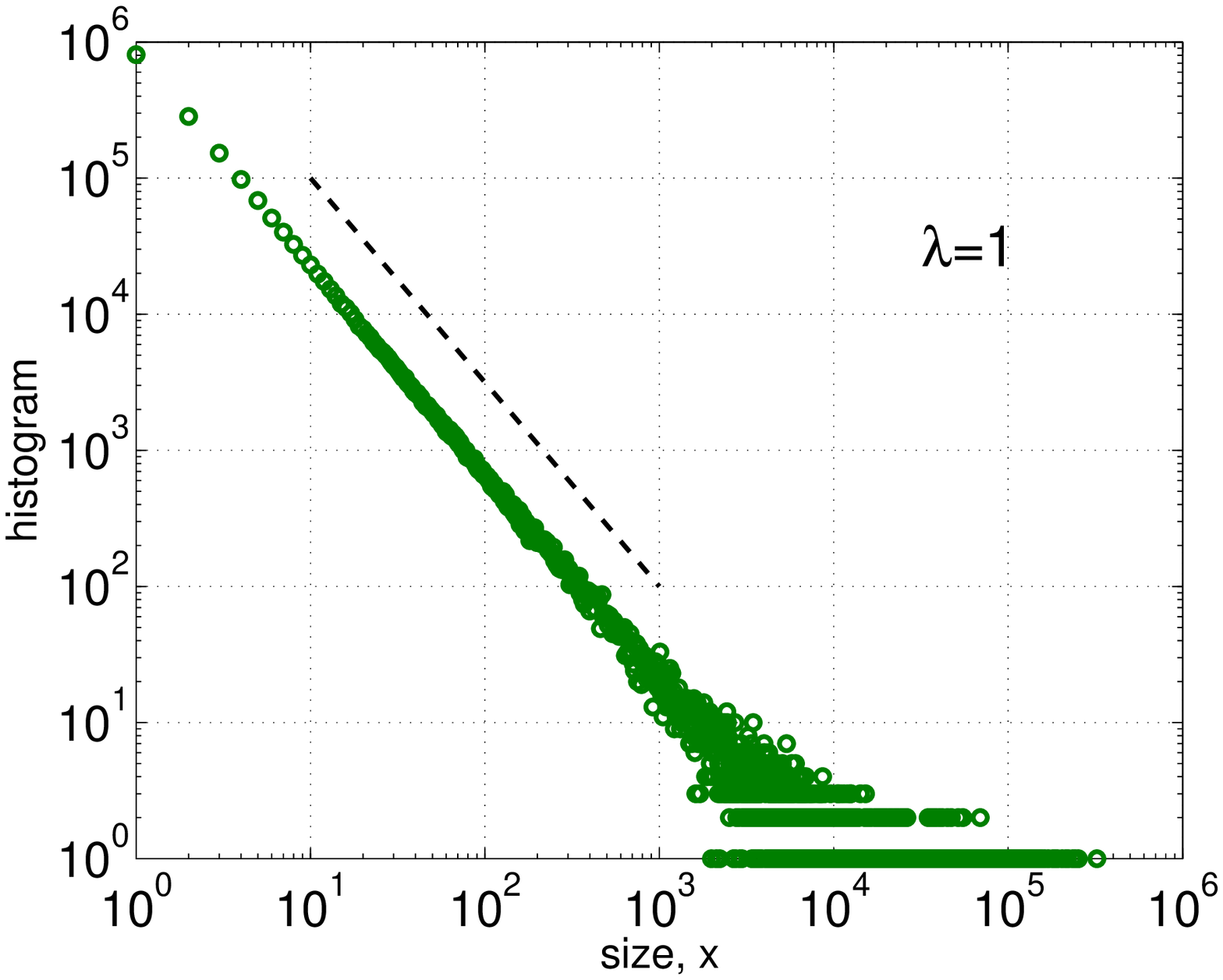, clip =,width=0.32\linewidth }
	\epsfig{file=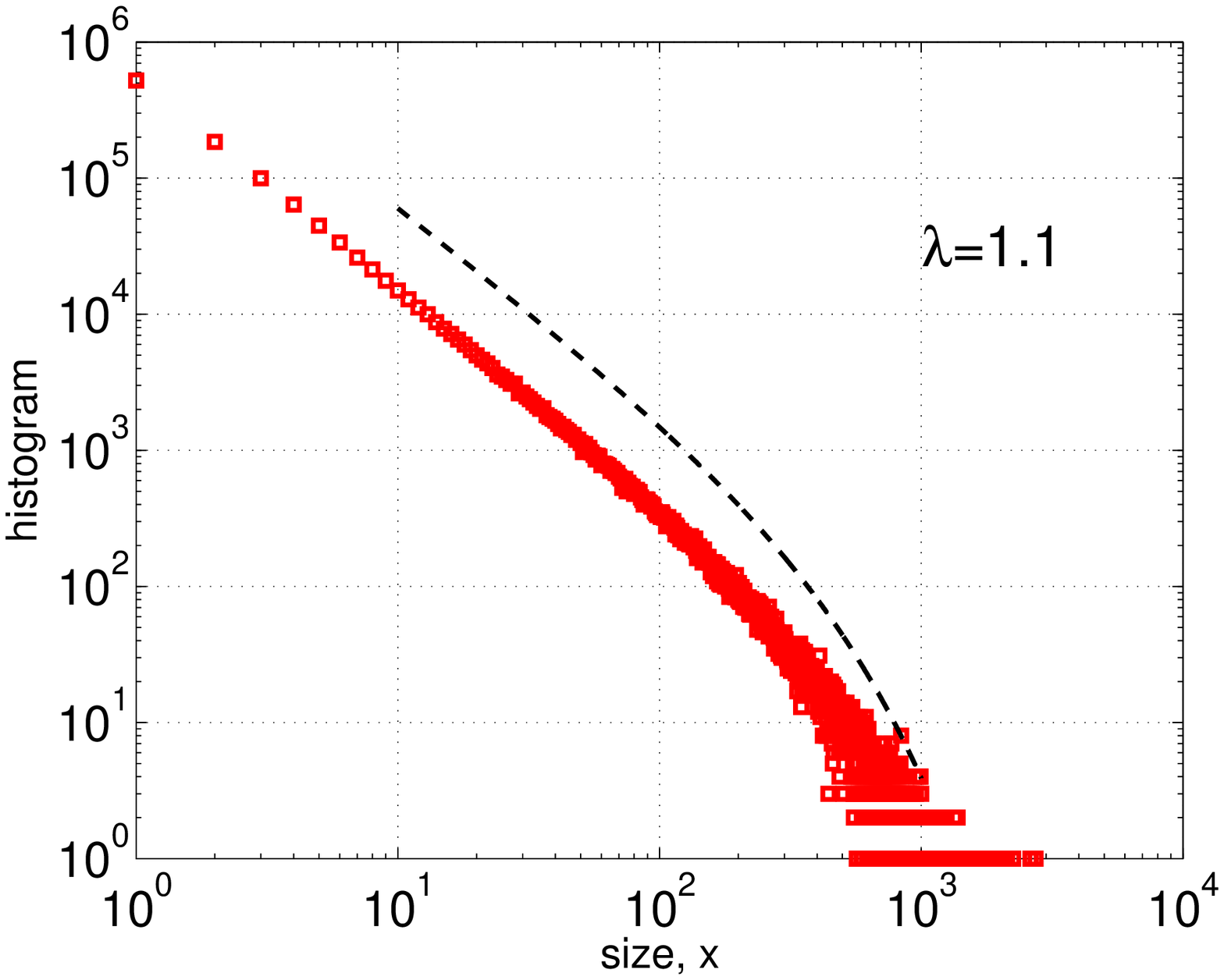, clip =,width=0.32\linewidth }
	\caption{(color online) Histograms of avalanche size shown above for networks of $N=10^5$ nodes with power-law degree distribution, exponent $\gamma = 3.5$ with Perron-Frobenius eigenvalues of $\lambda = 0.9$ (left), $\lambda=1.0$ (center) and $\lambda=1.1$ (right) on a log-log scale. Symbols show the number of avalanches having size $x$ from a single simulation of $10^6$, $2 \times 10^6$, and $10^6$ avalanches, respectively, from left to right. Dashed lines provide a reference for the theoretical prediction $x^{-3/2}\exp(-x/x^{*})$ described in Eqs.~\eqref{eq-plexponential} and \eqref{eq-xstar}. Note that the vertical position of the dashed lines was chosen arbitrarily. Infinite size avalanches in the supercritical case (right) are not represented in the data set. Agreement between theoretical prediction and measurement is excellent despite finite sample size noise.}
	\label{figure-size}
\end{figure*}
The asymptotic form for the distribution of the size of avalanches starting at node $n$, $p_n(x)$, can be obtained from the asymptotic form of $g_n(s)$ as $s\to 0$. Therefore, we study Eq.~\eqref{eq-gn} by  assuming $g_n(s)$ is small.  In order to obtain an analytic expression for the distribution of size we assume, in addition, that the network is close to critical, $|\lambda_D-1| \ll 1$. Taking logarithms in Eq.~\eqref{eq-gn} and using the approximation $\ln(1+g) \approx g - g^2/2$ we obtain
\begin{align}
	\label{eq-gnsD}
	g_n(s) - \frac{1}{2}g_n(s)^2 &=\nonumber \\
	 -s + \sum_{m=1}^{N} H_{mn}g_m(s) &- \frac{1}{2} \sum_{m=1}^{N} H_{mn}^2g_m^2(s).
\end{align}
As $s\rightarrow 0$ and $g_n \to 0$, the leading order terms are $g_n(s) = -s + \sum_m H_{mn} g_m(s)$, or $(H^T-I) {\bf g} = s {\bf 1}$, where ${\bf g} = [g_1,g_2,\dots,g_N]^T$ and ${\bf 1} = [1,1,\dots,1]^T$. When $|\lambda_D -1|\ll 1$, and $\lambda_D$ is well separated from the rest of the spectrum of $H$, as we are assuming, ${\bf g} = s (H^T-I)^{-1} {\bf 1} \sim {\bf v}$, where ${\bf v}$ is the left Perron-Frobenius eigenvalue of $H$ (more precisely, we are assuming such a separation for $A$, but since $H=A$ when $\lambda=1$ and we are assuming $|\lambda_{D} -1| \ll 1$, by continuity the assumption is valid for $H$ as well). Since ${\bf v}$ is independent of $s$, the solution up to first order is approximately $g_n(s) = g(s)v_n/\langle v\rangle$, where the term $\langle v\rangle = \frac{1}{N}\sum_{n=1}^N v_n$ is included to make $g(s)$ independent of the normalization of ${\bf v}$. For small $s$, and including the nonlinear terms, we expect the solution of Eq.~\eqref{eq-gnsD} to be close to this solution, so we set
\begin{equation}
	\label{eq-gsu}
	g_n(s) = g(s)\langle v\rangle ^{-1}v_n + \varepsilon_n(s),
\end{equation}
where $\varepsilon_n$ is a small unkown error term.  Substituting Eq.~\eqref{eq-gsu} into Eq.~\eqref{eq-gnsD}, using $H^T {\bf v} = \lambda_D {\bf v}$, and neglecting terms of order $\varepsilon g$ we get
\begin{align}
	&g(s)\langle v\rangle ^{-1}v_n+\varepsilon_n(s)-\frac{1}{2}g(s)^2\langle v\rangle ^{-2}v_n^2 \notag \\
	&= -s + \lambda g(s)\langle v\rangle ^{-1}v_n \notag \\
	&+ \sum_{m=1}^{N} H_{mn}\varepsilon_m(s) - g(s)^2\langle v\rangle ^{-2}\frac{1}{2}\sum_{m=1}^{N}H_{mn}^2v_m^2.
\end{align}
To eliminate the unknown error term $\varepsilon_n$, we multiply by the right eigenvector entry $u_n$ of $H$ and sum over $n$.  We use $H {\bf u} = \lambda_D {\bf u}$  and neglect $(\lambda_D - 1)\varepsilon_n$ to get
\begin{align}
\label{eq-vgs}
	&g(s)\langle v\rangle ^{-1}\langle uv\rangle -\frac{1}{2}g(s)^2\langle v\rangle ^{-2}\langle uv^2\rangle  \notag\\
	&= -s\langle u\rangle  + \lambda_D g(s)\langle v\rangle ^{-1}\langle uv\rangle\notag\\
	&-g(s)^2\langle v\rangle ^{-2}\frac{1}{2N}\sum_n \sum_m u_nH_{mn}^2v_m^2,
\end{align}
where $\langle x y\rangle  \equiv \frac{1}{N}\sum_nx_ny_n$.  Equation~\eqref{eq-vgs} is a quadratic equation for $g(s)$, $ag^2+bg+c=0$, with
\begin{align}
	a &= \frac{\sum_n\sum_mu_n(1-H_{mn}^2)v_m^2}{2N\langle uv\rangle \langle v\rangle }, \\
	b &= (\lambda_D - 1), \\
	c &= -s \frac{\langle u\rangle \langle v\rangle }{\langle uv\rangle }.
\end{align}
Solving for $g(s)$ and substituting back into $g_n(s) = \phi_n(s) - 1$ we find, choosing the root that guarantees $g_n < 0$,
\begin{equation}
	\phi_n(s) = 1 + \frac{-(\lambda_D - 1) - \sqrt{(\lambda_D-1)^2+4sa\frac{\langle u\rangle \langle v\rangle }{\langle uv\rangle }}}{2a} \frac{v_n}{\langle v\rangle}
	\label{eq-readyforinverse}
\end{equation}
The moment generating function $\phi_{n}$, first defined in Eq.~\eqref{mgf}, can be interpreted as the Laplace transform of the distribution of size. Taking the inverse Laplace transform of the form of $\phi_{n}(s)$ found in Eq.~\eqref{eq-readyforinverse} we obtain that for large $x$, the distribution of size $p_n(x)$ is approximately given by
\begin{equation}
	p_n(x) \propto  v_n x^{-3/2}\exp(-x/x^*),
	\label{eq-plexponential}
\end{equation}
where the characteristic size $x^*$ is given by
\begin{equation}
	x^* = 4a\frac{\langle v\rangle \langle u\rangle }{\langle vu\rangle } \frac{1}{(\lambda_D-1)^{2}}.
	\label{eq-xstar}
\end{equation}
The distribution of size is asymptotically an exponential times a power-law with exponent $-3/2$.  Such a functional form describes the distribution of the size of connected clusters near the percolation threshold in some network percolation models \cite{Newman2001,Cohen2002}. In the critical case, when $\lambda = \lambda_D = 1$, $x^*$ diverges and we recover a power-law distribution with exponent $-3/2$, which is the well-known exponent for critical branching processes \cite{Harris1963, Zapperi1995}.  It is interesting to note that this exponent, in our model, does not depend on the structure of the network, contrasting related percolation models where all nodes with the same degree are considered statistically equivalent \cite{Cohen2002}. Also note that the quantity $a$ in Eq.~\eqref{eq-xstar} depends implicitly on $\lambda_D$.

\section{Numerical Experiments}\label{section-experiments}
In this section, we test the theoretical predictions of the previous sections by directly simulating the process described in Sec.~\ref{section-formulation} on computer-generated networks.  We first describe the processes used to construct networks and simulate avalanches. 

Networks were constructed in two steps. First, binary networks (with adjacency matrix entries $\hat A_{mn} \in \{ 0,1\}$) were constructed via an implementation of the configuration model \cite{configurationModel}, using $N=10^5$ nodes, with nodal degrees drawn from a power-law distribution with exponent $3.5$, i.e., the probability that a node has degree $k$ is proportional to $k^{-3.5}$. Second, each nonzero entry $\hat A_{mn}$ was given a weight, drawn from a uniform distribution $\mathcal{U}[0,1]$. We then calculated the Perron-Frobenius eigenvalue of this weighted matrix, $\hat{\lambda}$, and multiplied the matrix by $\lambda / \hat{\lambda}$, resulting in a matrix $A$ with the desired eigenvalue $\lambda$. We simulated avalanches for networks with $\lambda$ between $0.5$ and $1.5$, sampling more finely for values close to $1$.   

Each simulated avalanche was created by first exciting a single network node, chosen uniformly at random, and then calculating the size and duration of the resulting avalanche as defined in Eqs.~\eqref{define-duration} and \eqref{define-size}. If the resulting avalanche lasted for more than $10^6$ time steps, we considered it as having infinite duration and infinite size. In all cases, the initial excitation was included so that the minimum size was $x=1$ and the minimum duration was $d=1$. For each subcritical ($\lambda < 1$) and supercritical ($\lambda > 1$) case, $10^6$ avalanches were simulated, and for $\lambda=1$, we simulated $2 \times 10^6$ avalanches to better sample the very broad distribution of avalanche size at criticality. 

A brief summary of the predictions of Secs.~\ref{section-duration} and \ref{section-size} is as follows. The probability of an avalanche of duration $d$ will decay as $\lambda^{d}$ for subcritical networks ($\lambda < 1$), as $d^{-2}$ for critical networks ($\lambda = 1$), and as $\lambda_{D}^{d}$ for supercritical networks ($\lambda > 1$), where $\lambda_D$ is the Perron-Frobenius eigenvalue of the matrix $D$, Eq.~\eqref{eq-h}. When $|\lambda_D -1|\ll 1$, the probability of a finite avalanche of size $x$ will decay as $x^{-3/2}\exp(-x/x^{*})$, where $x^{*}$ is a network-specific constant, given in Eq.~\eqref{eq-xstar}.
\begin{figure}[b]
	\centering
	\epsfig{file=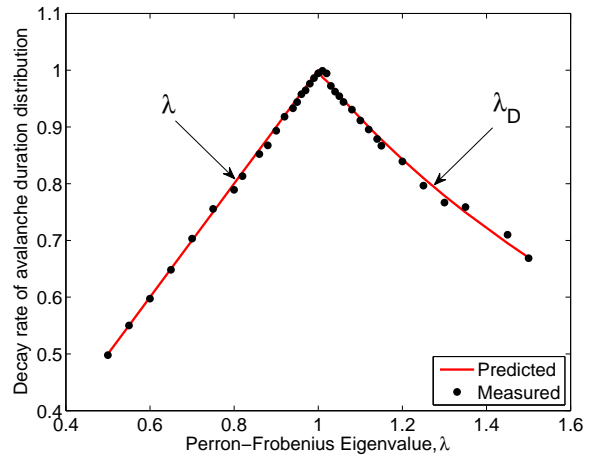, clip =,width=1.0\linewidth }
	\caption{(Color online) A comparison of predicted duration decay rates [Eq.~\eqref{eq-expApproach1} and Eq.~\eqref{eq-supercriticaldurationprediction}] (solid line), and numerical simulations (solid circles) plotted against $\lambda$, the largest eigenvalue of the network adjacency matrix. Agreement is excellent for both the subcritical and supercritical numerical simulations. The distribution of avalanches durations decays as $\lambda^{t}$ and $\lambda_{D}^{t}$ for $\lambda  \leq 1$ and $\lambda > 1$, respectively, as indicated by arrows.}
	\label{figure-durationagreement}
\end{figure}

\begin{figure}[t]
	\centering 
	\epsfig{file=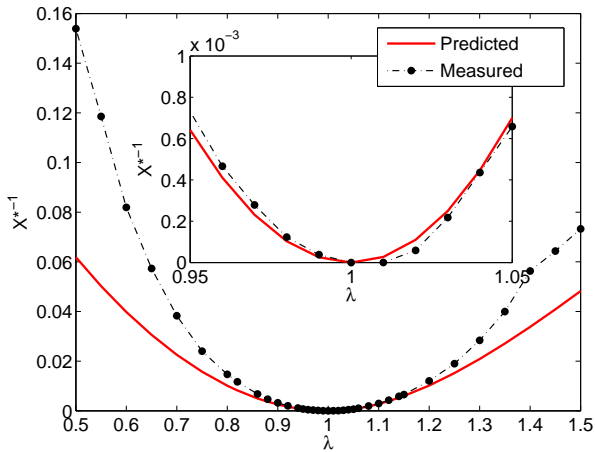, clip =,width=1.0\linewidth }
	\caption{(color online) Testing the prediction that avalanche size $x$ is distributed as $x^{-3/2}\exp{(-x/x^{*})}$, we compare the theoretical prediction of $x^{*}$ (solid line) with $x^{*}$ estimated via regression on the largest 10\% of avalanches from numerical simulations (solid circles, dashed line). Inset, identical data on a magnified domain around $\lambda=1$. Agreement is excellent for $\lambda$ near 1, and decreasingly accurate for much larger or smaller $\lambda$.}
	\label{figure-sizeagreement}
\end{figure}

In Figs.~\ref{figure-duration} and \ref{figure-size}  we compare histograms of avalanche duration and size obtained from direct numerical simulations for $\lambda=0.9$ (left), $1.0$ (center), and $1.1$ (right) with the theoretical predictions described in the previous paragraph (dashed lines). Note that, since our predictions allow for an unspecified proportionality constant, the vertical position of the dashed lines was chosen arbitrarily. In general, we find good agreement between the theoretical predictions of avalanche duration and size distributions with the histograms observed in the simulations. While the dashed lines in Figs.~\ref{figure-duration} and \ref{figure-size} are appealing to the eye, more quantitative measures of agreement between theory and experiment are shown in Figs.~\ref{figure-durationagreement} and \ref{figure-sizeagreement}. 

To numerically test the agreement between theory and experiment for the distribution of avalanche duration, in Fig.~\ref{figure-durationagreement} we compare the best fit $\hat \lambda$ of the data to $p(t) \propto \hat \lambda^t$, calculated through a nonlinear least-squares exponential regression on the simulated PDF of avalanche duration, to our theoretical predictions in Eqs.~\eqref{pdf1} and \eqref{pdf2} (solid line). The agreement is excellent, though not exact, over the entire range of $\lambda$ values simulated.

As a partial test of our theory for the distribution of avalanche size, we assume that the form of the distribution is $x^{-3/2}\exp(-x/x^{*})$, and estimate $x^{*}$ from the data, which we then compare with our theoretical prediction in Eq.~\eqref{eq-xstar}. Noting that as $\lambda \to 1$, $x^{*}$ will diverge, we estimated $1/x^{*}$ via a nonlinear least-squares using Brent's minimization on the cumulative histogram of the avalanche size data. Since our theory describes only the asymptotic form of the distribution, this estimate was performed only on the largest $10\%$ of measured data. [Similar results were obtained using the largest 5\%, 1\% and 0.1\% of data (not shown), but when using more than the largest 10\%, the minimizing $x^{*}$ value diverged, suggesting that we fit the power-law portion of data at the expense of the exponential tail.] Figure~\ref{figure-sizeagreement} shows the theoretical prediction (solid line) and the result of the numerical fit to the data (solid circles; the dashed lines are to aid the eye). As shown, agreement is quite good close to $\lambda_D =1$ (see the inset of Fig.~\ref{figure-sizeagreement}), but less accurate for very subcritical or supercritical networks. The latter is reasonable since the assumption that $|\lambda_D -1|$ is a small quantity was used in the derivation of Eq.~\eqref{eq-xstar}.

\begin{figure}[t]
	\centering 
	\epsfig{file=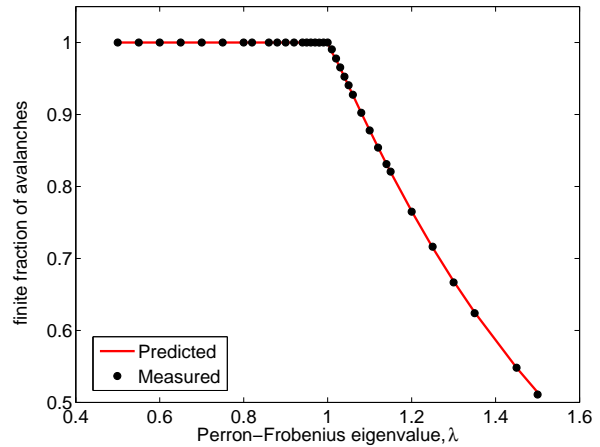, clip =,width=1.0\linewidth }
	\caption{When the Perron-Frobenius eigenvalue $\lambda$ is larger than one, there is a non-zero probability of an avalanche starting at node $n$ having infinite duration, as predicted by Eq.~\eqref{eq-bnFixedPoint}. Here we average the finite fraction of avalanches originating from node $n$ over all nodes, showing excellent agreement between the fraction predicted by averaging Eq.~\eqref{eq-bnFixedPoint} (solid line) and fraction measured from simulation (solid circles).}
	\label{figure-finitefraction}
\end{figure}

\begin{figure}[t]
	\centering 
	\epsfig{file=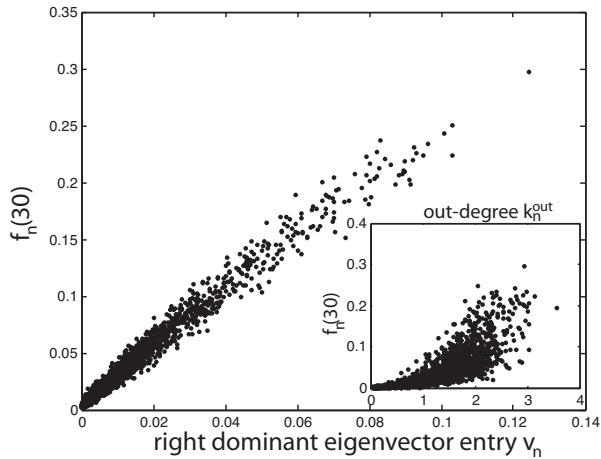, clip =,width=1.0\linewidth }
	\caption{Testing the node-specific prediction of Eq.~\eqref{eq-expApproach1}, avalanches were simulated on a subcritical ($\lambda = 0.95$) and disassortative ($\rho = 0.8$) Erd\H os R\'enyi random network with $N=10^4$ nodes.  In the large plot, the fraction of avalanches originating at node $n$ that last longer than 30 time steps, $f_n(30)$, is plotted against the corresponding entry in the right Perron-Frobenius eigenvector, $v_n$. In the inset, the same values $f_n(30)$ are plotted against the corresponding out-degree $k_{n}^{out}$. The eigenvector entry $v_n$ does a significantly better job than out-degree $k_{n}^{out}$ of predicting the duration of avalanches originating at node $n$ in disassortative networks (shown) and for assortative networks (not shown). }
	\label{figure-assortativity}
\end{figure}

Although Figs.~\ref{figure-durationagreement} and \ref{figure-sizeagreement} demonstrate agreement between theory and measurement for supercritical networks, that analysis was restricted to finite avalanches. To complement this result, we compare the predicted fraction of infinite avalanches with the measured fraction, for various values of $\lambda_D$. The quantity $b_n$ in Eq.~\eqref{eq-bnFixedPoint} is the fraction of avalanches originating at node $n$ which will have finite duration and size. In Fig.~\ref{figure-finitefraction}, we show the fraction of avalanches that decay in finite time, averaged over nodes, comparing theory (solid line)  with experiment (solid circles).  The theoretical fraction of avalanches was calculated by numerically solving Eq.~\eqref{eq-bnFixedPoint} to find $b_n, n = 1\dots, N$, and then plotting $\sum_{n=1}^N b_n /N$ as a function of $\lambda$.  The numerical  fraction of finite avalanches was calculated by simulating $10^6$ avalanches, each one starting at a random node (out of $N=10^5$ nodes). If an avalanche lasted more than $10^6$ steps, we counted it as an infinite avalanche. Then, an estimate of $b_n$ was calculated as the fraction of finite avalanches starting at node $n$. The symbols in Fig.~\ref{figure-finitefraction} show $\sum_{n=1}^N {b}_n /N$ as a function of $\lambda$. 
Agreement is excellent over the entire range of $\lambda$ values tested.
Beyond aggregate statistics, we also test a more subtle prediction of Eq.~\eqref{eq-expApproach1}. In Sec.~\ref{section-duration}, we concluded that $f_n(t) = 1-c_n(t)$, the probability that an avalanche started at node $n$ lasts more than $t$ steps, scales for large $t$ as $f_n(t) \propto \lambda^t v_n$, where ${\bf v}$ is the left Perron-Frobenius eigenvector of $A$. Other research in the network adjacency matrix literature has noted that the vector of nodal out-degrees (in-degrees) is a good approximation for the right (left) dominant eigenvector of $A$ in the absence of degree correlations \cite{eigenvalueApproximation}. In this light, our prediction above is understandable: when there are not degree correlations in the network, a node with a larger right eigenvector entry (and thus larger out-degree) will tend to produce longer avalanches. Therefore, in order to fully test our prediction, we created networks with {\it assortative mixing by degree} \cite{newmanAssortativity}, a type of degree correlation which we measure using the coefficient $\rho$ \cite{eigenvalueApproximation},
\begin{equation}
	\rho = \frac{\langle k_n^{in}k_m^{out}\rangle _e}{\langle k_n^{in}\rangle _e\langle k_m^{out}\rangle _e},
	\label{eq-rho}
\end{equation}
where $\langle \cdot \rangle_{e}$ denotes an average over all edges and $k$ are weighted nodal degrees defined as $k_n^{in} = \sum_m A_{mn}$ and $k_n^{out} = \sum_m A_{nm}$. In the absence of degree correlations between connected nodes $\langle k^{in}_{n} k^{out}_{m} \rangle_{e} = \langle k^{in}_{n} \rangle_{e} \langle k^{out}_{m} \rangle_{e}$ and $\rho = 1$. In assortative networks, there exists a positive correlation ($\rho > 1$) between the in-degree at node $n$ and the out-degree at node $m$ at the ends of a directed link from $n$ to $m$. When the correlation is negative ($\rho < 1$), the network is called disassortative. Thus we created Erd\H os-R\'enyi random networks with $N=10^4$ nodes, and rewired each network via a link-swapping process (as described in Ref.~\cite{eigenvalueApproximation}) until we had very assortative and disassortative networks ($\rho = 1.2$ and $\rho = 0.8$, respectively). Eq.~\eqref{eq-expApproach1} implies that in such networks, the tails of the CDF of avalanches originating at node $n$ will be proportional to the corresponding entry of the right eigenvector, which may differ significantly from the nodal out-degree. For the a subcritical network ($\lambda = 0.95$) with assortativity coefficient $\rho = 0.8$ we plot $f_n(30)$ and its corresponding entry in the right dominant eigenvector $v_n$ for each node $n$, in Fig.~\ref{figure-assortativity}, showing that proportionality is excellent. In the inset of the same figure we plot $f_n(30)$ against the corresponding out-degree $k^{out}_n$ for each node $n$, showing that proportionality to out-degree does not hold. Assortative networks produce the same effect, but are not shown here.

\section{Discussion}\label{section-discussion}

We have presented an analysis of the asymptotic distributions of the duration and size of avalanches in complex networks. This work is of interest in various applications, most notably neuroscience \cite{Petermann2009,Shew2009,Stewart2008,Kinouchi2006,Larremore2011-PRL,Larremore2011-Chaos,Tanaka2009,Beggs2003,Shew2011,Benayoun2010} and the analysis of power-grid failure cascades \cite{Dobson}. While some of our results, such as the functional forms for the distributions, are analogous to those found in classical Galton-Watson branching processes \cite{Harris1963} or in mean-field models \cite{Cohen2002}, we emphasize the distinguishing aspects of our results: (i) We generalize the criterion for criticality to $\lambda = 1$, which depends on the topology of the network in ways that previous results do not capture. For example, in critical branching processes \cite{Zapperi1995} the condition for criticality is $\langle d \rangle = 1$. (ii) The parameters of the asymptotic distributions in the various regimes are affected by the network topology, and our results allow us to predict how various factors such as network degree distributions or degree-degree correlations affect these parameters [e.g., the parameter $x^*$ in Eq.~\eqref{eq-xstar} or $\lambda_D$ in Eq.~\eqref{eq-supercriticaldurationprediction}]. (iii) In contrast to previous studies, our results allow us to predict the statistics of avalanches generated at a particular node. This might be of critical importance in certain applications where the adjacency matrix is known or can be inferred (such as the power grid or the Autonomous System network of the internet) since one can then allocate resources to prevent avalanches, if so desired, that start at the nodes which tend to generate the largest avalanches. As shown in Fig.~\ref{figure-assortativity}, the naive prediction that the nodes with the largest out-degree generate the largest avalanches is not necessarily true when the networks have nontrivial structure, such as degree correlations. 

In developing our theory, we made some assumptions which we now discuss. First, we assumed that the network was locally tree-like. This allowed us to treat avalanches propagating to the neighbors of a given node as independent of each other. While this is a good approximation for the networks we used, it is certainly not true in general. In particular, avalanches propagating separately from a given node might excite the same node as they grow. The result is that the number of nodes that the avalanches excite in the simulation may be less than what the theory would predict. In running our simulations, we addressed this issue in two ways: first, we kept track of the number of times two branches of the same avalanche simultaneously excited the same node $n$, finding it to be an increasing function of avalanche size and Perron-Frobenius eigenvalue, yet still negligible when compared to the total number of excitations. In addition, each time such an event occurred, we separately generated an avalanche starting from the doubly excited node $n$ and corrected both the size and duration of the original avalanche by incorporating these additional avalanches. We found that doing this had no appreciable effect on the measured distributions, and so all figures shown in this manuscript are produced from simulation data {\it without} the additional compensating avalanches included. This, and the fact that the numerical simulations are described well by the theory, suggest that the interaction of avalanches propagating to different neighbor nodes can be safely neglected in the networks studied. The performance of our theory in networks that are not locally tree-like, such as networks with a high degree of clustering, is left for future research. Another approximation we used is that the Perron-Frobenius eigenvalue $\lambda$ is well separated from the rest of the spectrum. This is a good approximation in networks without well defined communities, but can break down in networks with strong community structure \cite{Chauhan2009}. 

Finally, we note that our results show that the experimental signatures of criticality in neural systems (characterized by a power-law distribution of avalanche size and duration with exponents $-3/2$ and $-2$, respectively \cite{Beggs2003,Shew2009,Petermann2009,Shew2011}) are robust to complex underlying network topologies.

The authors acknowledge useful discussions with Woodrow Shew. DBL and MC were supported by NSF MCTP Grant No DMS-0602284. JGR was supported by NSF Grant No. DMS-090822. EO was supported by ONR MURI Gran N00014-07-1-0734.

\appendix

\section{Probability of finite avalanche duration}\label{appendixa}

In this Appendix we establish that the probability of finite avalanches, under our assumptions, is always one when $\lambda \leq 1$ (critical and subcritical networks), and becomes less than one when $\lambda > 1$ (supercritical networks). These probabilities, $b_n = \lim_{t\to \infty}c_n(t)$, satisfy the equation
\begin{equation}
	\label{bns}
	b_n = \prod_{m=1}^N \Bigl[ (1-A_{mn}) + A_{mn}b_m \Bigr].
\end{equation}
First, we show that if  $\lambda \leq 1$, where $\lambda$ is the Perron-Frobenius eigenvalue of $A$, then the only solution to the equation above is $b_n = 1$. Letting $b_n = 1 - f_n$, we have for all $n$
\begin{equation}
	\label{fns}
	1-f_n = \prod_{m=1}^N \Bigl[ 1 - A_{mn}f_m \Bigr].
\end{equation}
Using the Weierstrass product inequality \cite{Weierstrass},
\begin{equation}
	\label{fns2}
	\sum_{m=1}^NA_{mn}f_m \geq 1 - \prod_{m=1}^{N} \Bigl[ 1 - A_{mn}f_m \Bigr] = f_n,
\end{equation}
with equality only if (i) $A_{mn} f_m = 0$ for all $m$, or (ii) $A_{mn} f_m = 0$ for all $m \neq k$ and $A_{kn} f_k = 1$ for some $k$ \cite{Weierstrass}. If $u$ is the right Perron-Frobenius eigenvector of $A$, this implies, since ${\bf u}^T A^T = \lambda {\bf u}^T$,
\begin{equation}\label{lala}
{\bf u}^T A^T {\bf f} =  \lambda {\bf u}^T  {\bf f} \geq {\bf u}^T  {\bf f}.
\end{equation}
If there is a nonzero $f_n$, then ${\bf u}^T  {\bf f} > 0$ since the Perron-Frobenius eigenvector has positive entries for irreducible $A$. Therefore, if $\lambda < 1$ we must have $f_n = 0$ for all $n$. If $\lambda = 1$, Eq.~\eqref{lala} implies equality in Eq.~\eqref{fns2}, which implies either (i) $A_{mn}f_m = 0$ for all $m$, and thus  $f_n = 0$ by \eqref{fns2}, or (ii) $A_{mn} f_m = 0$ for all $m \neq k$ and $A_{kn} f_k = 1$ for some $k$, which is impossible since we assumed that the entries of $A$ are strictly less than one and $f_k$ is a probability. Therefore, we must have $f_n = 0$ if $\lambda = 1$, a valid argument for any $n$. Together with the previous argument above, we conclude that $b_n = 1$ for all $n$ if $\lambda \leq 1$.

Now, we show that if $\lambda > 1$ then $\lim_{t\to \infty} c_n(t) = b_n < 1$. To show this, we view Eq.~\eqref{eq-durationupdate} as a dynamical system, and note that the analysis of Sec.~\ref{subcritical}, applied to the case $\lambda > 1$, shows that the fixed point $b_n = 1$ is linearly unstable. If we show that $c_n(t)$ is nondecreasing with $t$, then the limit $b_n$ must be less than one. By induction, we will prove that $c_n(t+1) \geq c_n(t)$ for all $n$. First, we have $c_n(0) = 0$ and $c_n(1) = \prod_m (1-A_{nm}) \geq 0$, so the statement is valid for $t = 0$. Then, assume $c_m(t) \geq c_m(t-1)$ for all $m$ and consider $c_n(t+1)/c_n(t)$, noting that $c_n(t) > 0$:

\begin{align}\label{eq-durationupdateproof}
\frac{c_n(t+1)}{c_n(t)} &= \prod_{m=1}^N \frac{(1-A_{mn}) + A_{mn}c_m(t)}{(1-A_{mn}) + A_{mn}c_m(t-1)}.\nonumber \\
 &= \prod_{m=1}^N \left[1 + \frac{ A_{mn}(c_m(t) - c_m(t-1))}{(1-A_{mn}) + A_{mn}c_m(t-1)}\right] \nonumber \\
 & \geq 1,
\end{align}
which proves the desired statement. Note that, although from the definition \eqref{eq-definecn}, it follows that $c_n(t)$ are nondecreasing, this proof is necessary since Eq.~\eqref{eq-durationupdate} is an approximation.

\section{$\lambda > 1 \Rightarrow \lambda_D  < 1$}\label{appendixb}

In this Appendix we argue that the Perron-Frobenius eigenvalue of the similar matrices $H$ and $D$ is less than one when the Perron-Frobenius eigenvalue of $A$ is greater than one: $\lambda > 1 \Rightarrow \lambda_D < 1$. Recall that the matrix $D$ was defined as
\begin{equation}\label{h}
D_{mn} = \frac{b_nA_{mn}}{(1-A_{mn})+b_mA_{mn}},
\end{equation}
where $b_n$, the probability that an avalanche starting at node $n$ is finite, satisfies
\begin{equation}
\label{bn}
b_n = \prod_{m=1}^N \Bigl[ (1-A_{mn}) + A_{mn}b_m \Bigr].
\end{equation}

Now, suppose that $A$ is such that $\lambda > 1$, and introduce a parameter $\alpha \leq 1$ by defining $b_n(\alpha)$ as the $b_n$ corresponding to the matrix $\alpha A$, which satisfies
\begin{equation}
b_n(\alpha) = \prod_{m=1}^N \Bigl[ (1-\alpha A_{mn}) + \alpha A_{mn}b_m(\alpha) \Bigr].
\end{equation}
Now, calculate the derivative of $b_n(\alpha)$ with respect to $\alpha$,
\begin{equation}
\frac{db_n(\alpha)}{d\alpha} = b_n(\alpha) \sum_{m=1}^N \frac{- A_{mn} +  A_{mn}b_m(\alpha) + \alpha A_{mn} \frac{db_m(\alpha)}{d\alpha} }{(1-\alpha A_{mn}) + \alpha A_{mn}b_m(\alpha)}.
\end{equation}
Letting $\mu_n = \frac{db_n}{d\alpha}\big|_{\alpha = 1}$, and evaluating the expression above at $\alpha = 1$, we get
\begin{align}
\mu_n =& b_n \sum_{m=1}^N \frac{- A_{mn} +  A_{mn}b_m + A_{mn} \mu_m }{(1- A_{mn}) +  A_{mn}b_m}\\
=& \sum_{m=1}^N D_{mn}(b_m-1) + \sum_{m=1}^N D_{mn}\mu_m.
\end{align}
In matrix form,
\begin{align}
(D^T-I){\bf \mu} = D^T({\bf 1} - {\bf b}),
\end{align}
where ${\bf 1} = [1,1,\dots,1]^T$, ${\bf b} = [b_1,b_2,\dots,b_N]^T$, and ${\bf \mu} = [\mu_1,\mu_2,\dots,\mu_N]^T$. Now, we left multiply the previous equation by ${\bf u}^T$, where ${\bf u}$ is the right Perron-Frobenius eigenvector of $D$, satisfying ${\bf u}^T D^T = \lambda_D {\bf u}^T$, to get
\begin{align}
	\label{eq-beeeight}
	(\lambda_D-1){\bf u}^T{\bf \mu} = \lambda {\bf u}^T({\bf 1} - {\bf b}).
\end{align}
If $\lambda > 1$, Appendix \ref{appendixa} shows that the entries of $({\bf 1} - {\bf b})$ are all positive. Since the Perron-Frobenius eigenvector ${\bf u}$ has positive entries as well (since we are assuming $A$ is irreducible), the right hand side of Eq.~\eqref{eq-beeeight} is positive.  Now, we argue that the vector ${\bf \mu}$ has nonpositive entries: as $\alpha$ increases, the probability of an excitation passing between any pair of nodes increases, and thus the probability of having a finite avalanche can not increase, i.e., $db_n/d\alpha \leq 0$. Therefore, the term ${\bf u}^T{\bf \mu}$ on the left hand side must be nonpositive and, since the right hand side is nonzero, it must be negative. Thus, the term $\lambda_D -1$ must be negative, that is, $\lambda_D < 1$.

\bibliographystyle{plain}

\begin{thebibliography}{99}
\bibitem{Weierstrass} The Weierstrass product inequality, which can be proved by induction on $N$, states that if $c_n\in [0,1]$ for $n = 1,2,\dots,N$, then $\prod_{n=1}^N (1-c_n) + \sum_{n=1}^N c_n\geq 1$, with equality only if all the $c_n$ are zero, or all the $c_n$ are zero except one of them, which is equal to one.
\end{thebibliography}

\end{document}